# Black phosphorus mode-locked Er-doped ZBLAN fiber laser at 2.8 μm wavelength

Zhipeng Qin,[1] Guoqiang Xie,[1,3] Chujun Zhao,[2] Shuangchun Wen,[2] Peng Yuan,[1] and Liejia Qian[1,4]

[1]Key Laboratory for Laser Plasmas (Ministry of Education), Department of Physics and Astronomy, Collaborative Innovation Center of IFSA (CICIFSA), Shanghai Jiao Tong University, Shanghai 200240, China
[2]Key Laboratory for Micro-/Nano-Optoelectronic Devices of Ministry of Education, Collaborative Innovation Center of IFSA (CICIFSA), School of Physics and Electronics, Hunan University, Changsha 410082, China
[3]email: xiegq@sjtu.edu.cn
[4]email: qianlj19@sjtu.edu.cn

**Abstract:** Mid-infrared saturable absorber mirror is successfully fabricated by transferring the mechanically exfoliated black phosphorus onto the gold-coated mirror. With the as-prepared black phosphorus saturable absorber mirror, a continuous-wave passively mode-locked Er:ZBLAN fiber laser is demonstrated at the wavelength of 2.8 μm, which delivers a maximum average output power of 613 mW, a repetition rate of 24 MHz and a pulse duration of 42 ps. To the best of our knowledge, it is the first time to demonstrate black phosphorus mode-locked laser at 2.8 μm wavelength. Our results demonstrate the feasibility of black phosphorus flake as a new two-dimensional material for application in mid-infrared ultrafast photonics.

Two-dimensional (2D) materials, represented by graphene, topological insulators (TIs) and transition metal dichalcogenides (TMDCs) [1-6], have aroused considerable attention in recent years due to their broadband absorption, ultrafast carrier dynamics and planar characteristic [7-8]. They have been regarded as the potential candidate for next-generation optoelectronic devices such as photoelectric detector, field-effect transistor, optical modulator, and so on [9-11]. Few-layer black phosphorus (BP), a new kind of 2D material, joined the family in 2014 [12].

BP crystal is a layered material, in which individual atomic layers are stacked together by van der Waals interaction. Inside a single layer, each phosphorus atom is covalently bonded with three adjacent phosphorus atoms to form a puckered honeycomb structure [13]. The three bonds take up all the valence electrons of phosphorus, so monolayer BP is a semiconductor which has a direct bandgap of ~2 eV. For multilayer BP, the bandgap decreases with the increasing thickness due to the interlayer interactions, and eventually down to ~0.3 eV for bulk BP [14]. No matter in monolayer or multilayer BP, they have a direct bandgap which significantly benefits to optoelectronic application. Multilayer BP can be cost-effectively fabricated by the liquid phase exfoliation [15-21] or mechanical exfoliation method [22-26]. So far in fiber lasers, the BP Q-switched lasers have been realized at the wavelength of 1.6, 1.9 and 2.8 μm [16, 20, 27], and the BP mode-locked lasers have been reported at the wavelength of 1.6 and 1.9 μm [17, 23-26]. In all-solid-state lasers, BP Q-switched lasers have been realized at the wavelength of 0.6, 1.0, 2.1 and 2.7 μm [19, 21, 22], and BP mode-locked laser is only demonstrated at 1 μm [18]. However, so far there is no report on BP mode-locked laser at the wavelength of 3 μm no matter in fiber lasers or in all-solid-state lasers. It is noticed that 3-μm mode-locked fiber lasers have been realized by employing saturable absorbers (SA) such as Fe:ZnSe, InAs and SESAM [28-31], but these SAs need complex fabrication process and high cost in comparison with BP. The traditional nonlinear polarization evolution technology can be also used for mode locking of 3-μm fiber laser [32, 33], but it requires complicated cavity structure and more mid-infrared optical components.

In this Letter, we demonstrated a BP mode-locked Er-doped ZBLAN fiber laser at the wavelength of 2.8 μm. A scotch tape-based mechanical exfoliation method was used to peel thin BP flakes from bulk BP crystal (XF Nano, INC) and then BP flakes was transferred onto the gold-coated mirror as BP saturable absorber mirror (BP SAM). By employing the as-prepared BP SAM, the continuous-wave (CW) mode-locked Er:ZBLAN fiber laser was realized, which delivered an average power of 613 mW and pulse duration of 42 ps at a repetition rate of 24 MHz. Our research results show that BP flake is a promising SA for mid-infrared pulsed lasers.

Firstly, the BP flake was transferred onto $CaF_2$ wafer for characterization. Stimulated by 532 nm laser, the Raman spectrum of the BP flake is shown in Fig. 1a. Three peaks located at 361 $cm^{-1}$, 437 $cm^{-1}$ and 465 $cm^{-1}$ can be clearly observed from the Raman spectrum, which correspond to $A_g^1$, $B_{2g}$ and $A_g^2$ vibration modes of BP crystal lattice, respectively. The $B_{2g}$ and $A_g^2$ modes arise from the in-plane oscillation of phosphorus atoms layer, and $A_g^1$ mode relates to out-of-plane vibration of phosphorus atoms layer [34]. The scanning electron microscopy (SEM) image of BP flakes with an amplification rate of 1000 is shown in Fig. 1b. It can be seen that the mechanically-exfoliated BP flake is uniform with high quality. In order to confirm the thickness of the as-prepared BP flake, we performed the thickness measurement with atomic force microscopy (AFM), as shown in Fig. 1c and 1d. It is worth noticing that some bubble-like bulge exists in the BP flakes, which results in height fluctuation in Fig. 1d. According to the height difference between A and B position, the BP flake has a thickness of ~143 nm. In consideration of monolayer thickness of 0.6 nm for BP, the as-prepared BP flake is estimated to be of 238 layers. According to the BP bandgap equation $E_g \approx (1.7/n^{0.73}+0.3)$ eV ( n is the number of layers) [14], our BP flakes has a bandgap of 0.33 eV, which means the as-prepared BP flakes can operate up to 3.8 μm wavelength.

We transferred the BP flake onto the gold-coated mirror to serve as BP SAM. The goad-coated mirror had a reflectivity of 99% at 2.8 μm. After combination with BP flake, the nonlinear reflection was measured by employing a home-made mode-locked fiber laser which delivered a maximum average power of 1.05 W and a repetition rate of 22 MHz at 2.8 μm. The measurement results are shown in Fig. 2. The reflectivity of BP SAM increases with the incident pulse fluence, which clearly exhibits the saturable absorption characteristic of BP SAM. From the trace, the modulation depth and saturable fluence of BP SAM are 19% and 9 μJ/$cm^2$, respectively.

The schematic of the mode-locked laser is shown in Fig. 3. In the linear cavity configuration, a highly Er-doped ZBLAN fiber was pumped by a commercial 976-nm fiber-coupled laser diode (BTW, INC), which delivered a maximum power of 30 W with a fiber core diameter of 105 µm and a numerical aperture (NA) of 0.15. The pump beam was collimated and then focused by two antireflectively coated lenses ($f_1$=50 mm, $f_2$=100 mm, T=99.5% @ 976 nm). The 6 mol.% Er-doped double clad ZBLAN fiber (Fiber Labs, INC) has a core diameter of 30 µm, a NA of 0.12 and a length of 4.2 m. The 1st cladding configuration has a diameter of 300 µm and a NA of 0.51, which allows for efficient pump coupling. The 2nd cladding diameter of the fiber is 425 µm. The front end of the fiber was perpendicular-cleaved, which served as output coupler with Fresnel reflection (4%) as feedback. The 45 ° placed quartz mirror (T>95% @ 976 nm, R>99% @ 2800 nm) was used to separate the laser beam from the pump. The rear end of the fiber was cleaved at an angle of 8 °to avoid parasitic oscillation. Both ends of the fiber were covered by indium foil and mounted in an aluminum semicircular groove for efficient heat removal. In the cavity, two highly-reflective concave mirrors (R=99.7%@ 2800 nm) were used to reimage the end facet of fiber onto BP SAM. Unlike lens-based imaging system [29], this concave mirror-based imaging system can avoid the intracavity feedback, which is beneficial to obtain stable mode-locking.

At the pump power of 1.1 W, the laser operated in CW regime. Then it switched to Q-Switching and Q-switched mode-locking (QML) regime when the incident pump power was beyond 1.7 W. When the pump power reached up to 3.3 W, stable CW mode-locking (CWML) was realized. A maximum average output power of 613 mW was achieved under an incident pump power of 5.35 W. At the maximum output power, we monitored the power stability of the laser by power meter. The output power fluctuation was only ± 0.2 % and the CWML operation could be sustained for more than half an hour. At the maximum output power, we didn't observe any damage of BP SAM. When we further increased the incident pump power beyond 5.35 W, the CWML operation became instable due to thermal instability of BP flake [18].

The mode-locked pulse train was captured by an infrared HgCdTe optoelectronic detector with a rise time of < 2 ns (VIGO System model PCI-9), and displayed in a digital oscilloscope with 500-MHz bandwidth (Tektronix, DPO3054). Figure 5a shows the typical mode-locked pulse trains in nanosecond and microsecond time scales. The radio frequency (RF) spectrum of the mode-locked pulses was measured by the RF spectrum analyzer (Agilent E4402B), as shown in Fig. 5b. Both the pulse trains and RF spectrum were measured at the maximum output power. The pulse repetition rate (i.e., the fundamental RF peak) is 24.27 MHz, corresponding to the effective cavity length of 6.2 m. The signal-to-noise ratio (SNR) in RF spectrum reaches to 60 dB, and no other radio frequency component is observed, which shows a clear CWML operation. The RF spectrum with a wider span from 10 to 250 MHz is also given in the inset of Fig. 5b.

At maximum average output power, autocorrelation trace of the output pulses was measured using a commercial autocorrelator (APE pulse Check USB MIR) with a time window of 150 ps, as shown in Fig. 6a. Assuming a Gaussian pulse profile, the mode-locked pulse duration is 42 ps. The corresponding mode-locked pulse spectrum was recorded by an optical spectrum analyzer (Ocean Optics, SIR 5000) with a resolution of 0.22 nm, as shown in Fig. 6b. The spectrum has a full width at half-maximum (FWHM) of 2.8 nm centered at 2783 nm. The time bandwidth product is calculated to be 4.5, which indicates the mode-locked pulses are chirped.

In conclusion, we have demonstrated a BP mode-locked Er-doped ZBLAN fiber laser at 2.8 µm for the first time. The mode-locked Er:ZBLAN fiber laser has generated pulses with an average output power of 613 mW (pulse energy of 25.5 nJ), repetition rate of 24 MHz, and pulse duration of 42 ps. Our research results suggest that BP is a promising SA in mid-IR spectral regime. In addition, the advantages of low cost, easy fabrication and thickness-dependent variable bandgap make BP more attractive as the next-generation optoelectronics devices.

This work is partially supported by Shanghai Excellent Academic Leader Project (Grant No. 15XD1502100), National Basic Research Program of China (Grant No. 2013CBA01505), and National Natural Science Foundation of China (Grant No. 11421064).

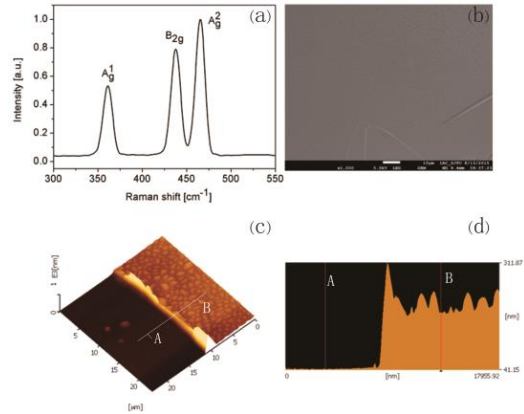

Fig. 1. (a) Raman spectrum of BP flake sample. (b) Morphology of BP flake measured by the scanning electron microscopy (SEM). (c) Three dimensional morphology of BP flake scanned by atomic force microscopy (AFM). (d) Thickness of BP flake.

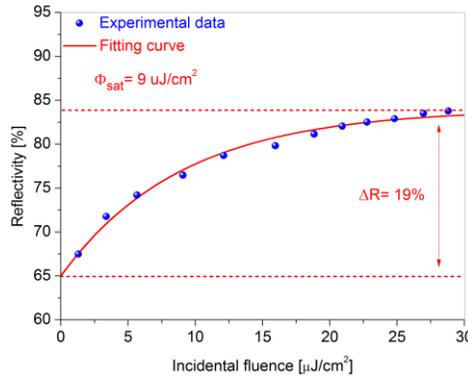

Fig. 2. The saturable absorption measurement of BP SAM at 2.8 μm.

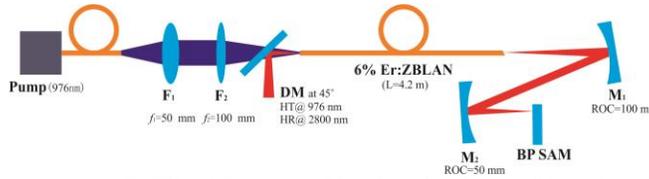

Fig. 3. Schematic of the mode-locked Er:ZBLAN fiber laser. DM, dichroic mirror; ROC, radius of curvature; BP SAM, black phosphorus saturable absorber mirror.

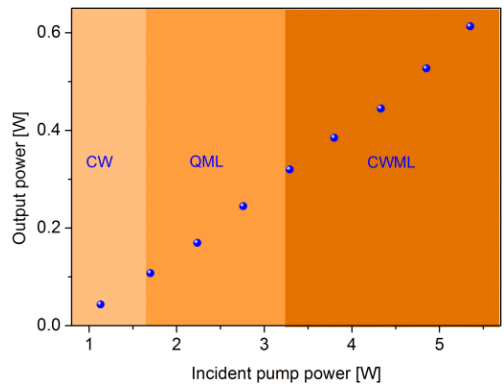

Fig. 4. Output average power versus incident pump power in different regimes. CW, continuous-wave; QML, Q-switched mode-locking; CWML, continuous-wave mode-locking.

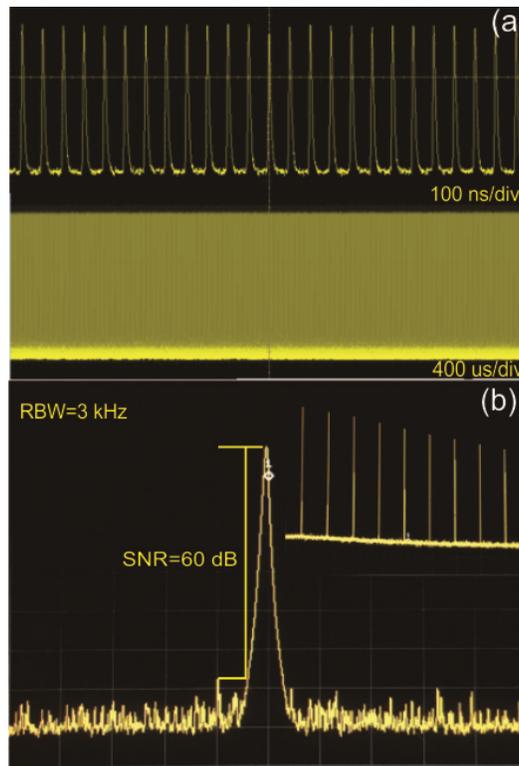

Fig. 5. (a) CW mode-locked pulse trains in nanosecond and microsecond time scales. (b) RF spectrum of the mode-locked pulse trains with a span of 0.6 MHz. Inset: the wide-span RF spectrum from 10 to 250 MHz.

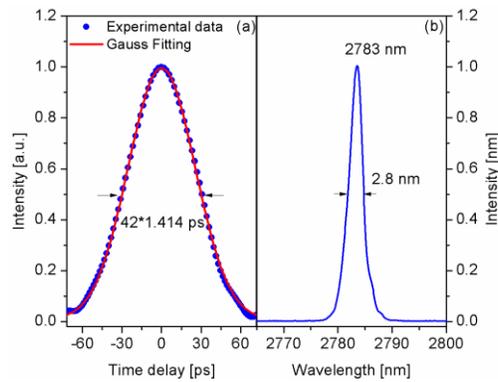

Fig. 6. (a) Autocorrelation trace of the mode-locked pulses. (b) The corresponding mode-locked pulse spectrum